# When do supernova neutrinos of different flavors have similar luminosities but different spectra?

H.-Thomas Janka*

Department of Astronomy and Astrophysics,

University of Chicago,

5640 S. Ellis Avenue, Chicago, IL 60637, U.S.A.

March 16, 1995

## Abstract

Muon and tau neutrinos ($\nu_x$) interact with protoneutron star matter only via neutral currents and exchange energy with the stellar gas predominantly by neutrino-electron scattering and neutrino-pair processes. In contrast, electron neutrinos and antineutrinos ($\nu_e$ and $\bar\nu_e$) are frequently absorbed and produced in charged-current mediated reactions with nucleons. Therefore the emergent $\nu_e$ and $\bar\nu_e$ originate from layers with lower temperatures further out in the star and are emitted with much lower characteristic spectral temperatures. In addition, a major contribution to the $\nu_e$ and $\bar\nu_e$ opacities is due to absorptions, while the opacity of $\nu_x$ is strongly dominated by scattering reactions with nucleons and nuclei in which the $\nu_x$ energy is (essentially) conserved. Therefore the $\nu_x$ distribution is nearly isotropic when $\nu_x$ decouple energetically and their outward diffusion is slowed down. In a generalized form to include this effect, the Stefan-Boltzmann Radiation Law can account for both the facts that $\nu_e$ ($\bar\nu_e$) and $\nu_x$ emerge from the star with similar luminosities but with very different characteristic spectral temperatures. Simple analytical expressions to estimate the effect are given. If, as recently argued, even at densities significantly below nuclear matter density neutral-current scatterings were associated with considerable energy transfer between neutrino and target particle, one might expect spectral temperatures of $\nu_x$ much closer to those of $\nu_e$ and $\bar\nu_e$. This is of relevance for the detection of neutrino signals from supernovae.

---

*On leave from Max-Planck-Institut für Astrophysik, Karl-Schwarzschild-Str. 1, D-85740 Garching, Germany.

# 1 Introduction

The transport of $\nu_e$ and $\bar{\nu}_e$ on the one hand and of heavy-lepton neutrinos $\nu_\mu$, $\bar{\nu}_\mu$, $\nu_\tau$, and $\bar{\nu}_\tau$ on the other is different in protoneutron star matter. Heavy-lepton neutrinos can be handled as one species $\nu_x$, because for vanishing or small (of the order of few eV) neutrino masses differences of their behavior are insignificant. Muons and tau mesons are not present in the outer regions of protoneutron stars; the maximum temperatures $T$ in newly formed neutron stars are several 10 MeV only and at densities $\rho \lesssim \rho_{\text{nuc}}$ ($\rho_{\text{nuc}} \approx 2.5 \cdot 10^{14}\,\text{g/cm}^3$ is the nuclear matter density) the electron chemical potential is usually below 100 MeV, too low to allow for a significant production of muons. Therefore electron-type neutrinos interact via neutral and charged currents with the particles of the stellar gas, while heavy-lepton neutrinos can react only through neutral currents. This difference affects neutrino-electron scattering and neutrino-pair processes, which are the most important reactions by which $\nu_x$ exchange energy with the stellar gas. The main effect on the transport, however, results from absorptions of $\nu_e$ and $\bar{\nu}_e$ onto $n$ and $p$, respectively. These reactions (and the inverse neutrino-producing processes) enforce reactive and thermal equilibrium between $\nu_e$ and $\bar{\nu}_e$ and the stellar gas until further out in the star, to lower densities and temperatures, where the weaker $\nu\bar{\nu}$-pair production processes are already frozen out for all neutrinos and where $\nu_x$ are therefore energetically decoupled from the background.

$\nu_e$, $\bar{\nu}_e$, and $\nu_x$ with the same energy scatter off $n$ or $p$ with the same cross sections. Since neutrino-electron scattering and neutrino-pair annihilation yield only a minor contribution ($< 10\%$) to the total transport opacity, the latter is dominated by neutrino-nucleon scatterings in case of $\nu_x$. In contrast, neutrino absorptions onto baryons provide a sizable fraction ($\approx$ 30–80%) of the transport opacity in case of $\nu_e$ and $\bar{\nu}_e$. This leads to different characteristics of the particle transport through the semi-transparent outer stellar layers, where the spectral distributions of the emitted neutrinos are determined.

$\nu_e$ and $\bar{\nu}_e$ emerge from regions with lower temperatures further out in the star and are emitted with much smaller characteristic spectral temperatures than $\nu_x$. Nevertheless, detailed models and transport simulations [1] show that all neutrino types are emitted with very similar luminosities. Unless $\nu_e$ and $\bar{\nu}_e$ emission from accretion onto the protoneutron star plays an important role the luminosities are equal to much better than a factor of 2. (For a review of the model results and more details, see [2].) These findings will be discussed and explained by the different transport properties of $\nu_e$ and $\nu_x$ in the presented paper. Simple analytical relations will be presented that allow one to estimate the size of the considered effect (Sect. 6). The implications and consequences will be addressed in Sect. 7.

# 2 The neutrino "spheres"

The differences of the interactions as described above lead to different relative locations of the "energy spheres" and "transport spheres" for neutrinos of different types. The "energy sphere" is defined as the layer in the star external to which a neutrino of representative energy, before leaving the star, interacts only once by a reaction with energy transfer to or from the target particle (optical depth for reactions with energy exchange $\approx 1$). The transport sphere is the corresponding layer with optical depth about unity for *all* reactions, including those which leave the neutrino energy (essentially) unchanged ("conservative processes").

Since a significant fraction of the total transport opacity of $\nu_e$ and $\bar{\nu}_e$ is due to absorptions onto nucleons, this process, however, is absent in case of $\nu_x$, the energy and transport spheres of $\nu_e$ and $\bar{\nu}_e$ lie very close together and further out at lower temperature and density. For $\nu_x$ both spheres are spatially clearly separated by the strong dominance of conservative scattering reactions in the transport opacity. Typically, from inside out the $\nu_x$ energy sphere, the $\nu_x$ scattering sphere, the $\nu_e$ ($\bar{\nu}_e$) energy sphere, and the $\nu_e$ ($\bar{\nu}_e$) scattering sphere follow in the direction of decreasing temperature and density. Dependent on the ratio of the mean spectral energies of $\nu_e$ and $\bar{\nu}_e$ relative to $\nu_x$ and caused by the variation of absorption and scattering cross sections with roughly the square of the neutrino energy, the radial locations of $\nu_x$ scattering sphere and $\nu_e$ ($\bar{\nu}_e$) energy sphere may also follow in inverse order. For steep gradients of $T$ and $\rho$ at the protoneutron star surface, the differences of the radii may be small, while the density $\rho$ and temperature $T$ change appreciably. Initially, the transport character of $\bar{\nu}_e$ is very similar to that of $\nu_e$. As the deleptonization goes on and the protoneutron star matter gets increasingly neutron-rich, however, it becomes more and more similar to that of $\nu_x$.

# 3 The "paradox"

As the protoneutron star deleptonizes and the electron chemical potential $\mu_e$ and the electron fraction $Y_e$ decrease, the chemical potential of electron neutrinos approaches zero: $\mu_{\nu_e} = \mu_e + \mu_p - \mu_n \to 0$. By this time the total energy fluxes (luminosities) of all neutrinos ($\nu_e$, $\bar{\nu}_e$, and $\nu_x$) become very similar in all detailed numerical models, $L_{\nu_e} \approx L_{\bar{\nu}_e} \approx L_{\nu_x}$. Nevertheless, $\nu_e$ (and $\bar{\nu}_e$) leave the star with significantly lower average energies than $\nu_x$, typically $\langle \epsilon_{\nu_x} \rangle \approx (2 \ldots 2.5) \cdot \langle \epsilon_{\nu_e} \rangle \approx (1.3 \ldots 1.7) \cdot \langle \epsilon_{\bar{\nu}_e} \rangle$.

To state the paradox in the above situation more clearly, let us assume that neutrinos stay roughly in thermodynamical equilibrium with the stellar gas until they decouple energetically at their corresponding energy spheres (narrow decoupling layers) at radii $R_{\epsilon,\nu_i}$ ($\nu_i = \nu_e, \bar{\nu}_e, \nu_x$). This means that to zeroth order of accuracy the neutrino spectra can be represented by Fermi-Dirac distributions according to local equilibrium, the neutrino temperature being equal to the local gas temperature, $T_{\nu_i} = T(R_{\epsilon,\nu_i})$.

Moreover, since at "late" times the chemical potential of electron neutrinos is small in a large part of the protoneutron star, $\mu_{\nu_e} = -\mu_{\bar{\nu}_e} \approx 0$, and $\mu_{\nu_x} = 0$ is true for the $\nu_x$ chemical potential at all times, the emission of neutrinos from their corresponding energy spheres can roughly be considered as blackbody-like with $\langle \epsilon_{\nu_i} \rangle \propto T(R_{\epsilon,\nu_i})$. $\nu_x$ decouple at a somewhat smaller radius, but higher temperature, which is reflected by $T(R_{\epsilon,\nu_x})/T(R_{\epsilon,\nu_e}) \gtrsim 2$. From the Stefan-Boltzmann Radiation Law one calculates the total neutrino luminosity according to $L_{\nu_i} \propto R_{\epsilon,\nu_i}^2 T^4(R_{\epsilon,\nu_i})$. Since differences of the energy-sphere radii of different neutrinos are very small compared to the neutron star radius (because of the steep density and temperature gradients close to the surface), i.e. $R_{\epsilon,\nu_e} \approx R_{\epsilon,\bar{\nu}_e} \approx R_{\epsilon,\nu_x} \approx R_{\mathrm{ns}}$, one ends up with the conclusion that $L_{\nu_x} \gtrsim 16 \cdot L_{\nu_e}$. This is in strong contradiction to the results of self-consistent model calculations according to which the luminosities are equal within a factor of less than about 2. How can this be understood?

## 4  Protoneutron star cooling

The emergent neutrino flux is certainly not a perfect equilibrium emission from a well defined neutrino sphere but is much more complex with a non-equilibrium spectral shape and with neutrinos of different energies decoupling from the stellar gas at different radii and at different thermodynamical conditions, simply because the neutrino opacity varies roughly with the square of the neutrino energy. This might serve as an "explanation" of the described puzzle, but does not actually yield insight into the direct reason why the neutrino luminosities can be very similar despite of the spectral dissimilarities. In fact, there is a simple way to understand qualitatively the physical mechanism which is responsible for the observed result.

Firstly, note that in the surface layers of the protoneutron star, even somewhat inside of the neutrino energy spheres, the total neutrino energy flux (as the sum of the luminosities of all neutrinos and with general relativistic redshift being disregarded) is conserved and given by the energy-loss rate of the dense core region of the star. External to the corresponding neutrino-energy sphere the total energy flux of each individual neutrino type is constant. The heat capacity of the dense core being very large, it is the energy loss from the central part of the star that determines the Kelvin-Helmholtz time scale of neutrino cooling. During the Kelvin-Helmholtz phase the diffusion of neutrinos out of the very opaque regions governs the evolution. Due to the long diffusion time scales (order of seconds) the cooling proceeds quasi-stationarily with very slow changes of thermodynamical quantities and neutrino emission parameters. In contrast, the surface layers heat up or cool down quickly when energy is deposited or lost via neutrinos. Within a time scale of only several 10 ms the temperature and density structure re-adjusts by expansion or contraction to assure that the neutrino-energy flow diffusing out from deeper regions is just transported through the surface layers.

Note, secondly, that for analogous reasons there is a conservation of the total *net* lepton-number fluxes (total number flux of neutrinos minus antineutrinos of each flavor) in the surface layers, but *no* conservation of the total number fluxes of individual neutrinos. Under the constraints that the total net lepton-number flux is conserved and the Pauli exclusion principle is obeyed, neutrinos are absorbed and re-emitted and redistributed in phase space. This is completely analogous to photons when they diffuse to the surface of a star with conserved total energy flux but with a downgrading in energy space as the temperature decreases towards the surface. The constant total energy flux is transported by an increasing number of neutrinos or photons. The redistribution in energy space determines the spectrum with which neutrinos are finally emitted from the neutrino-energy sphere, external to which energy changing interactions become very rare.

## 5  Fundamentals of radiative transfer

Considering neutrino diffusion through the background medium the neutrino energy flux is given by

$$\mathbf{F} \;=\; -D \cdot \nabla \left[ \frac{4\pi}{(hc)^3} T^4 \, \mathcal{F}_3(0) \right] \;=\; -\frac{\lambda_\mathrm{t} c}{3} \cdot 4 \cdot \frac{4\pi}{(hc)^3} \, \mathcal{F}_3(0) \, T^3 \nabla T \;, \tag{1}$$

where $D$ is the diffusion constant, $D = \lambda_\mathrm{t} c/3$, with the total transport mean free path $\lambda_\mathrm{t}$ (as determined from *all* neutrino reactions). $T$ is the gas temperature. Since neutrinos are assumed to be in thermal equilibrium with the stellar gas the considered transport approximation is called "equilibrium diffusion". $\mathcal{F}_3(0)$ denotes the third Fermi integral for relativistic particles with degeneracy parameter $\eta = \mu/T = 0$, the general definition being $\mathcal{F}_n(\eta) \equiv \int_0^\infty \mathrm{d}x \, x^n \left[ 1 + \exp(x - \eta) \right]^{-1}$. One can express the energy flux in terms of the "flux factor" $\mathbf{f}$ and the neutrino energy density $E$ by

$$\mathbf{F} \;=\; \mathbf{f} \, c \, E \;=\; \mathbf{f} \, c \, \frac{4\pi}{(hc)^3} T^4 \, \mathcal{F}_3(0) \;, \tag{2}$$

when $\mathbf{f}$ is defined as the normalized flux,

$$\mathbf{f} \;\equiv\; \int_{4\pi} \mathrm{d}\Omega \, \mathbf{\Omega} \, I(\mathbf{\Omega}) \Big/ \int_{4\pi} \mathrm{d}\Omega \, I(\mathbf{\Omega}) \;. \tag{3}$$

In spherical symmetry $f$ as the first angular moment of the radiation intensity $I(\mathbf{\Omega})$ gives the mean angle cosine of the radiation relative to the radial direction $\mathbf{r}/r \equiv \mathbf{n}$:

$$\mathbf{f} \;=\; f \cdot \mathbf{n} \;=\; \langle \cos\theta \rangle \cdot \mathbf{n} \;. \tag{4}$$

Combining Eq. (1) and Eq. (2) we find

$$\mathbf{f} \;=\; -\frac{4}{3} \cdot \lambda_\mathrm{t} \cdot \frac{\nabla T}{T} \;. \tag{5}$$

With Eq. (2) the neutrino luminosity of the protoneutron star can be expressed in terms of the conditions at $R_{\epsilon,\nu_i}$ as

$$\begin{aligned} L_{\nu_i} &= 4\pi R_{\epsilon,\nu_i}^2 \cdot f_{\nu_i}(R_{\epsilon,\nu_i}) \cdot c \, \frac{4\pi}{(hc)^3} \cdot T^4(R_{\epsilon,\nu_i}) \cdot \mathcal{F}_3(0) \\ &\propto R_{\epsilon,\nu_i}^2 \cdot f_{\nu_i}(R_{\epsilon,\nu_i}) \cdot T^4(R_{\epsilon,\nu_i}) \,. \end{aligned} \qquad (6)$$

If one assumes diffusion to be a suitable description until the energy flux decouples at the neutrino-energy sphere, $f_{\nu_i}(R_{\epsilon,\nu_i})$ is given by Eq. (5):

$$f_{\nu_i}(R_{\epsilon,\nu_i}) = -\frac{4}{3} \cdot \lambda_{t,\nu_i}(R_{\epsilon,\nu_i}) \cdot \left.\frac{\mathrm{d}\ln(T)}{\mathrm{d}r}\right|_{R_{\epsilon,\nu_i}} \,. \qquad (7)$$

In analogy to the photon case one can define an effective temperature $T_{\nu_i}^{\text{eff}}$ by

$$F_{\nu_i} = L_{\nu_i} \left/ \left(4\pi R_{\epsilon,\nu_i}^2\right)\right. = \sigma^* \left(T_{\nu_i}^{\text{eff}}\right)^4 \qquad (8)$$

with the radiation constant $\sigma^*$ for neutrinos given by

$$\sigma^* \equiv \frac{a^* c}{4} \,, \quad a^* \equiv \frac{4\pi}{(hc)^3} \cdot \mathcal{F}_3(0) \,. \qquad (9)$$

$a^*$ is the neutrino-density constant corresponding to the radiation-density constant $a$. The effective temperature $T_{\nu_i}^{\text{eff}}$ is related to the characteristic spectral temperature $T(R_{\epsilon,\nu_i})$ by comparison of Eq. (8) and Eq. (9) with Eq. (6):

$$T_{\nu_i}^{\text{eff}} = \left(4 f_{\nu_i}(R_{\epsilon,\nu_i})\right)^{1/4} \cdot T(R_{\epsilon,\nu_i}) \,. \qquad (10)$$

For causality reasons (i.e. $F \leq Ec$) and by Eq. (4) $f$ is physically limited to values $f \leq 1$. Close to and inside of the neutrino transport sphere one typically finds $f \lesssim 1/4$ [3, 4]. Therefore Eq. (10) implies that $T_{\nu_i}^{\text{eff}} \lesssim T(R_{\epsilon,\nu_i})$. Note that it is the characteristic spectral temperature $T_{\nu_i} \equiv T(R_{\epsilon,\nu_i})$ that is connected with the average energy $\langle \epsilon_{\nu_i} \rangle$ of the emitted neutrinos according to

$$\langle \epsilon_{\nu_i} \rangle = T_{\nu_i} \cdot \frac{\mathcal{F}_3(0)}{\mathcal{F}_2(0)} \,, \qquad (11)$$

if one assumes a vanishing spectral degeneracy parameter, $\eta_{\nu_i} = 0$.

## 6  The explanation

Eq. (6) and Eq. (7) allow us now to resolve the paradox of Sect. 3. From Eq. (6) one learns that the luminosity is not just proportional to the product $R_{\epsilon,\nu_i}^2 T^4(R_{\epsilon,\nu_i})$, but in a general treatment the flux factor $f(R_{\epsilon,\nu_i})$ appears. This is usually disregarded and $f(R_{\epsilon,\nu_i})$ is replaced by the value one obtains for a (black) spherical surface that radiates isotropically into vacuum:

$$f \;=\; \int_0^1 \mathrm{d}\mu\, \mu \Big/ \int_{-1}^1 \mathrm{d}\mu \;=\; \frac{1}{4} \qquad (12)$$

("vacuum approximation")[1]. The implicit assumption $f_{\nu_x}(R_{\epsilon,\nu_x}) = f_{\nu_e}(R_{\epsilon,\nu_e})$ is not justified in the case of $\nu_x$ and $\nu_e$ due to the differences of their transport behavior as discussed in Sects. 1 and 2. In detailed neutrino transport calculations with Monte Carlo methods [3, 4] values of $f_{\nu_i}(R_{\mathrm{t},\nu_i}) \approx 1/4$ are obtained for all types of neutrinos at their corresponding *transport* spheres but usually not at their energy spheres.

At the neutrino-energy spheres (index $\epsilon$) and transport spheres (index t) defined by optical depths of order unity, $\tau_{\mathrm{t},\epsilon} \approx 1$, the approximate relations hold:

$$\tau_{\mathrm{t},\epsilon} \;\approx\; \kappa_{\mathrm{t},\epsilon} \cdot R_{\mathrm{t},\epsilon} \;\approx\; 1 \quad \Longleftrightarrow \quad \lambda_{\mathrm{t},\epsilon} \;\approx\; R_{\mathrm{t},\epsilon} \;. \qquad (13)$$

$\lambda_{\mathrm{t}}$ and $\lambda_{\epsilon}$ are the transport mean free path and the mean free path for reactions with energy exchange, respectively, and $\kappa_{\mathrm{t},\epsilon} = \lambda_{\mathrm{t},\epsilon}^{-1}$ are the corresponding opacities[2]. The $\nu_x$ transport mean free path $\lambda_{\mathrm{t},\nu_x}$ is much smaller than the mean free path for energy exchange, because the opacity is dominated by conservative scattering reactions. Thus, $\lambda_{\mathrm{t},\nu_x}(r) \ll \lambda_{\epsilon,\nu_x}(r)$, therefore $\lambda_{\mathrm{t},\nu_x}(R_{\mathrm{t},\nu_x}) \approx R_{\mathrm{t},\nu_x} \approx R_{\mathrm{ns}}$ (Eq. (13)) and $\lambda_{\epsilon,\nu_x}(R_{\epsilon,\nu_x}) \approx R_{\epsilon,\nu_x} \approx R_{\mathrm{ns}}$, but $\lambda_{\mathrm{t},\nu_x}(R_{\epsilon,\nu_x}) \ll R_{\mathrm{ns}}$. In contrast, for $\nu_e$ (or $\bar\nu_e$) one has $\lambda_{\mathrm{t},\nu_e}(r) \approx \lambda_{\epsilon,\nu_e}(r)$ and therefore $\lambda_{\mathrm{t},\nu_e}(R_{\epsilon,\nu_e}) \approx R_{\mathrm{ns}}$. Putting all together one infers from Eq. (7)

$$f_{\nu_x}(R_{\epsilon,\nu_x}) \;\ll\; f_{\nu_e}(R_{\epsilon,\nu_e}) \;. \qquad (14)$$

Using this in Eq. (6) finally explains, why $L_{\nu_x} \approx L_{\nu_e}$ can be true although $T_{\nu_x} \gg T_{\nu_e}$. This is simply a consequence of the isotropization of the $\nu_x$ distribution due to frequent (iso-energetic) scatterings off nucleons, which slows down the $\nu_x$ diffusion and suppresses the $\nu_x$ flux from the $\nu_x$ energy sphere. Although $\nu_x$ decouple energetically deeper inside the protoneutron star and have a more energetic spectrum that reflects the higher temperature at the $\nu_x$ energy sphere, this leads to a $\nu_x$ luminosity similar to that of $\nu_e$.

To obtain a numerical estimate of the corresponding effect represented by Eq. (14), Eq. (7) can be evaluated for typical $\nu_x$ and $\nu_e$ energies with the relevant processes included. With the typical number $\langle\epsilon_{\nu_x}\rangle/\langle\epsilon_{\nu_e}\rangle \approx 2\ldots 2.5$ (Sect. 3) and with taking into account that the scattering opacity yields a fraction of 20%–70% of the total transport opacity of $\nu_e$ (or $\bar{\nu}_e$) (Sect. 1) and that the spectral averages of the scattering and absorption cross sections vary roughly with $\langle \epsilon^2 \rangle$, one estimates for the relation of the cross sections: $\sigma_{t,\nu_x}(R_{\epsilon,\nu_x}) \approx (1 \ldots 4) \cdot \sigma_{t,\nu_e}(R_{\epsilon,\nu_e})$. The cross sections become comparable when absorptions dominate scattering in the $\nu_e$ opacity. From this follows

$$\begin{aligned}
\frac{f_{\nu_x}(R_{\epsilon,\nu_x})}{f_{\nu_e}(R_{\epsilon,\nu_e})} &\approx \frac{\lambda_{t,\nu_x}(R_{\epsilon,\nu_x})}{\lambda_{t,\nu_e}(R_{\epsilon,\nu_e})} = \frac{\rho(R_{\epsilon,\nu_e})}{\rho(R_{\epsilon,\nu_x})} \frac{\sigma_{t,\nu_e}(R_{\epsilon,\nu_e})}{\sigma_{t,\nu_x}(R_{\epsilon,\nu_x})} \\
&\approx \left(\frac{1}{4} \ldots 1\right) \cdot \frac{\rho(R_{\epsilon,\nu_e})}{\rho(R_{\epsilon,\nu_x})} \; .
\end{aligned} \tag{15}$$

---

[1] Note that this is actually not a consistent definition, because one takes into account in the numerator only the part of the radiation that is directed outwards. For isotropic radiation the integration from $-1$ to $1$ would yield $f = 0$. Moreover, moving infinitesimally outward from the radiating surface one obtains $f = \frac{1}{2}$. Thus $f$ is discontinuous. For an introductory description in the context of photon transport, see, e.g., [5].

[2] Being precise, the determination of the neutrino-energy sphere requires the use of the *effective* mean free path for energy exchange, $\lambda_\epsilon^{\mathrm{eff}} = \sqrt{\lambda_\mathrm{t}\lambda_\epsilon}$, instead of $\lambda_\epsilon$. $\lambda_\epsilon^{\mathrm{eff}}$ represents the average displacement between two reactions with energy exchange (see, e.g., [5, 6, 3]). The presented arguments are qualitatively not affected by this.

Since $R_{\epsilon,\nu_x} < R_{\epsilon,\nu_e}$ one always has $\rho(R_{\epsilon,\nu_e})/\rho(R_{\epsilon,\nu_x}) < 1$ for the ratio of the densities at the $\nu_e$ and $\nu_x$ energy spheres. For steep density profiles the density ratio can become quite small, which is the case at late times during the protoneutron star cooling. At early times the neutronization of the star is not so strong and $\nu_e$ absorptions onto $n$ less frequent. Therefore the total transport cross section of $\nu_e$ and the ratio $\sigma_{t,\nu_e}(R_{\epsilon,\nu_e})/\sigma_{t,\nu_x}(R_{\epsilon,\nu_x})$ are smaller.

## 7 Discussion and implications

The result of detailed protoneutron star cooling models that the luminosities of emergent $\nu_x$ and $\nu_e$ ($\bar{\nu}_e$) are similar although the spectra are very different, can be understood as a consequence of the differences of the interaction of $\nu_x$ and $\nu_e$ ($\bar{\nu}_e$) with protoneutron star matter. $\nu_x$ interact by neutral currents only, while $\nu_e$ and $\bar{\nu}_e$ react via neutral and charged currents. The $\nu_x$ transport opacity is therefore dominated by (essentially) iso-energetic neutrino-$n$ and neutrino-$p$ scatterings, whereas absorptions onto nucleons yield a sizable contribution to the opacity of $\nu_e$ and $\bar{\nu}_e$. Due to the stronger coupling to the stellar matter $\nu_e$ and $\bar{\nu}_e$ emerge from the star further out, at lower temperatures and densities than $\nu_x$. All their interactions with the stellar gas cease at around the same radius. In contrast, when $\nu_x$ decouple energetically from the stellar gas at their energy sphere they still experience frequent scatterings off nucleons (and nuclei) on their way out. This "back-scattering" causes an isotropization of the $\nu_x$ distribution even exterior to the $\nu_x$ energy sphere and slows down the diffusion of $\nu_x$ to the stellar surface. The corresponding suppression of the $\nu_x$ luminosity leads to a significant discrepancy between the effective temperature $T_{\nu_x}^{\text{eff}}$ ($\propto L_{\nu_x}^{1/4}$) and the characteristic spectral temperature $T_{\nu_x}$ ($\propto \langle \epsilon_{\nu_x} \rangle$). The emergent $\nu_x$ spectrum remains nearly unchanged outside of the $\nu_x$ energy sphere and $T_{\nu_x}$ therefore reflects the temperature in a deeper layer of the star than $T_{\nu_e}$ or $T_{\bar{\nu}_e}$.

Instead of absorbing the effects of scattering-dominated $\nu_x$ transport into the effective temperature one can explicitly account for the different angular distributions of $\nu_e$ ($\bar{\nu}_e$) and $\nu_x$ by using a generalized form of the Stefan-Boltzmann Radiation Law that includes the flux factors $f_{\nu_i}(R_{\epsilon,\nu_i})$. Three parameters, $f_{\nu_i}(R_{\epsilon,\nu_i})$, $T_{\nu_i} = T(R_{\epsilon,\nu_i})$, and $R_{\epsilon,\nu_i} \approx R_{\mathrm{ns}}$ characterize the physical dependences of the luminosities. Considering the transport mean free paths of $\nu_e$ ($\bar\nu_e$) and $\nu_x$ at their corresponding energy spheres a simple analytical way to estimate the influence of the discussed effects on the properties of the emergent neutrino fluxes is given in Eq. (13)–Eq. (15).

A potential difference between the characteristic spectral temperature and the effective temperature according to Eq. (10) has been neglected in all evaluations of the SN 1987A neutrino signal that were based on simple cooling models of the nascent neutron star (e.g., [7]). In case of $\bar\nu_e$ this difference is quite pronounced at late times when the protoneutron star matter has become very $n$-rich and the $\bar\nu_e$ and $\nu_x$ transport are more similar than during the early phase of the cooling evolution. In time-dependent analyses of the neutrino data the effect can therefore not be ignored.

Deviations from an isotropic radiation field at the emitting surface account for the phenomenon of "limb darkening" which is strong when $f$ tends to values close to 1, indicating an outward peaked angular distribution of the radiation. In case of an extended, radiating surface the limb darkening effect in combination with the characteristic radiation temperature in principle allows one to determine the temperature structure as a function of depth below the surface (Eq. (7)). A reliable value of the factor $f(R_{\epsilon,\nu_i})$ is also crucial to recover neutron star radii from measured neutrino luminosities and characteristic spectral temperatures (Eq. (6); see also [2, 3]).

The addressed differences of $\nu_x$ and $\nu_e$ ($\bar\nu_e$) transport and the associated consequences for neutrino fluxes and spectra were based on the assumption that the energy exchange between neutrinos and target particles in neutrino-nucleon scatterings is negligible. However, as recently argued [8], this need not be true because neutrinos interact with nucleons in a dense plasma and not with isolated targets in a very dilute medium. For typical protoneutron star temperatures even at densities as low as $10^{13}$ g/cm$^3$ nucleon-nucleon collisions could allow for significant energy transfer between nucleons and neutrinos that is not possible in interactions with isolated nucleons due to the smallness of nucleon recoils. In the context of the presented paper this has the important implication that, if different neutrinos are not affected in the same way or not equally strongly, the differences between the spectra of $\nu_x$ and $\nu_e$ or $\bar\nu_e$ could be significantly changed. In the considered "standard" case of conservative neutrino-nucleon scatterings the neutrino-energy spheres, in particular the energy sphere of $\nu_x$, move inward to very high densities as the temperatures in the protoneutron star decrease along with the cooling process and the decrease of the neutrino opacities due to Pauli blocking in the Fermion phase spaces. Including the effects of nucleon-nucleon interactions in the dense medium might lead to more similar $\nu_x$ and $\nu_e$ spectra when the discrepancy between the $\nu_x$ transport mean free path and the mean free path for energy exchange shrinks. A quantitative analysis of this possibility needs more theoretical work on the cross sections; incorporation into the numerical modelling of the neutrino cooling of protoneutron stars might yield very interesting results. It is well possible that detections of neutrinos of all flavors from a future supernova, e.g. in the Superkamiokande and SNO experiments, might allow to determine the size of the discussed effect by a comparison of the $\nu_e$ and $\nu_x$ spectra.

**Acknowledgements.** This paper was inspired by conversations with M. Turner and G. Sigl. The author is very grateful to G. Raffelt for a reading of the manuscript and for a number of suggestions for improvement. This work was supported in part by the National Science Foundation under grant NSF AST 92-17969, by the National Aeronautics and Space Administration under grant NASA NAG 5-2081, and by an Otto Hahn Postdoctoral Scholarship of the Max-Planck-Society.